\documentclass[preprint,showpacs,amsmath,amssymb]{revtex4}

\usepackage{graphicx}
\usepackage{dcolumn}

\def\finesse{{\mathcal F}}

\begin{document}

\title{Enhanced photothermal displacement spectroscopy \\
	for thin-film characterization \\
	using a Fabry-Perot resonator}
	
\date{\today}
\author{Eric D. Black, Ivan S. Grudinin, Shanti R. Rao, Kenneth G. Libbrecht}
\affiliation{LIGO Project, California Institute of Technology \\
Mail Code 264-33, Pasadena CA 91125}

\begin{abstract}

We have developed a new technique for photothermal displacement spectroscopy that is potentially orders of magnitude more sensitive than conventional methods. We use a single Fabry-Perot resonator to enhance both the intensity of the pump beam and the sensitivity of the probe beam. The result is an enhancement of the response of the instrument by a factor proportional to the square of the finesse of the cavity over conventional interferometric measurements. 

In this paper we present a description of the technique, and we discuss how the properties of thin films can be deduced from the photothermal response. As an example of the technique, we report a measurement of the thermal properties of a multilayer dielectric mirror similar to those used in interferometric gravitational wave detectors.

\end{abstract}

\pacs{}

\maketitle

\newpage

\section{Introduction}

Recently there has been much interest in understanding noise mechanisms in dielectric coatings of mirrors used in advanced interferometric gravitational wave detectors~\cite{Fejer03, Braginsky03, Braginsky00, Nakagawa02, Yamamoto02-1, Yamamoto02-2, Harry02, Crooks02}. One noise mechanism that could potentially limit the sensitivity, and hence the astrophysical reach, of an advanced detector is coating thermoelastic-damping noise~\cite{Fejer03, Braginsky03, Braginsky99, Cerdonio01}. In order to accurately predict the level of this noise and to design coatings in which this noise source is minimized, we need to have accurate values for the thermal properties of the thin films that make up these dielectric coatings. Specifically, we need to know the thermal expansion coefficient $\alpha$ and the thermal conductivity $\kappa$ of the coating materials in thin-film form.

It is well known that the thermal conductivities of materials in thin films can differ markedly from those of the same materials in bulk form~\cite{Wu93, Grilli00, Wu97, Kuo92, Langer97}, and there is evidence of a similar deviation in the thermal expansion coefficient~\cite{Braginsky03-2, Inci02, Tien00}. Thus, there is a need to directly measure the thermal properties of candidate coatings in order to select ones that exhibit the lowest thermoelastic-damping noise.

We have constructed an apparatus to characterize dielectric coatings on mirrors for the purpose of selecting the best coating for an advanced interferometric gravitational wave detector. We use interferometric photothermal displacement spectroscopy, a technique well suited for measuring the thermal conductivity and expansion coefficient of a thin film~\cite{Olmstead83, Mandelis-books-1, Mandelis-books-2, Mandelis-books-3, Mandelis-books-4, Mandelis00}, but our method differs from conventional photothermal-displacement-spectroscopy techniques in one important point. We use a Fabry-Perot cavity to substantially enhance both the pump-beam heating power applied to the sample and the sensitivity of the probe beam. 

In this paper we both describe the apparatus and discuss a simple and intuitive way of looking at the physics of the photothermal response over a broad range of frequencies.

\section{The instrument}

Traditional interferometric photothermal displacement spectroscopy uses separate pump and probe beams, sometimes with different wavelengths, to locally heat a sample and measure the resulting thermal expansion. One widely-used configuration uses the sample as the end mirror in one arm of a Michelson interferometer (see for example~\cite{Olmstead83}). De Rosa \emph{et al.}~\cite{DeRosa02} have observed the differential photothermal response in a pair of Fabry-Perot cavities, in a configuration originally conceived as an optical readout for a resonant-bar gravitational-wave detector. Their experiment nicely demonstrated the photothermal effect in Fabry-Perot cavities, verifying the theory of Cerdonio \emph{et al.}~\cite{Cerdonio01} for the frequency-dependent photothermal response of a homogeneous material.

\begin{center}
\begin{figure}
\includegraphics{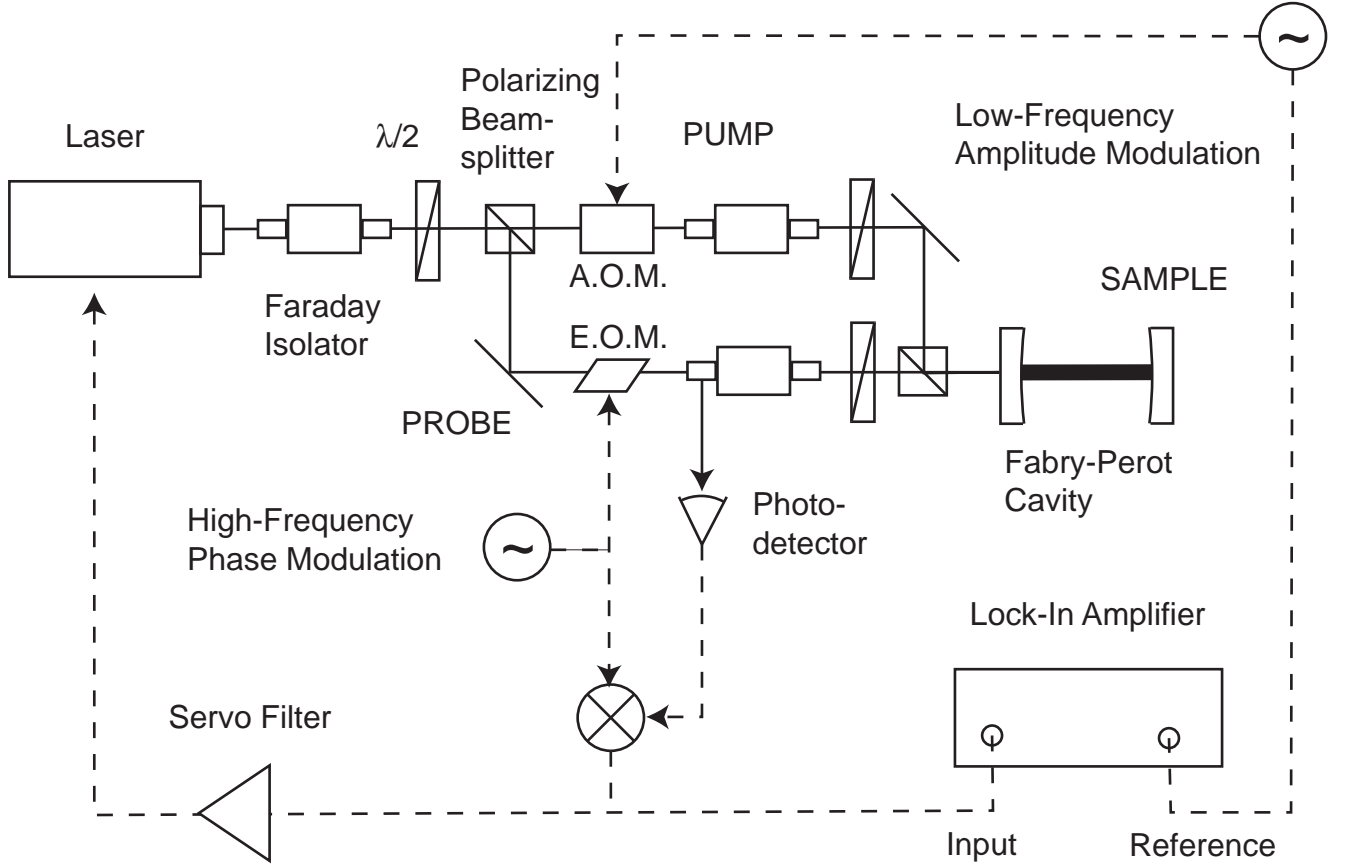}
\caption{ \label{fig:layout} A diagram of our experimental apparatus.}
\end{figure}
\end{center}

There is currently a need to characterize the thin-film dielectric coatings that are expected to be used in advanced interferometric gravitational wave detectors. We have the need to measure both the thermal expansion coefficient and thermal conductivity of such coatings, and so we have constructed an instrument based on interferometric photothermal displacement spectroscopy to perform such measurements on candidate coatings. Our apparatus uses a single Fabry-Perot cavity, as shown in Figure~\ref{fig:layout}, with the sample forming one of the mirrors in the cavity~\cite{Black00,Rao03}. 


Figure~\ref{fig:layout} shows a diagram of our experimental apparatus. Both the pump and probe beams are provided by the same laser, so that both can resonate simultaneously inside the cavity. The pump and probe beams are distinguished from each other by having orthogonal polarizations. The first half-wave plate after the laser adjusts the polarization, determining how much goes into the pump or probe paths via a polarizing beamsplitter. We phase modulate the probe beam using an electro-optic modulator (E.O.M.) and lock it (and by extension the pump beam) to the cavity by use of the Pound-Drever-Hall method~\cite{Drever83, Black01}. We amplitude-modulate the pump beam using an acousto-optic modulator (A.O.M.) and use lock-in detection to measure the photothermal response from the error signal. 

For a pump beam sinusoidally-modulated at a frequency $f$ that is much higher than the unity-gain frequency of the servo, the error signal is given by
\[
\varepsilon = -8 R P_{probe} \frac{L{\mathcal F}}{\lambda} J_0 (\beta) J_1 (\beta) \left\{ \frac{\delta L}{L} \sin 2 \pi f t \right\},
\]
where $R$ is the (RF) response of the photodiode, $P_{probe}$ is the power in the probe beam, $L$ and ${\mathcal F}$ are the length and finesse of the Fabry-Perot cavity, $\lambda$ is the wavelength of the laser light used, $\beta$ is the phase modulation depth (in radians),  and $\delta L \sin (2 \pi f t)$ is the deviation in $L$ from the ideal resonance condition $2L=N\lambda$. $J_0$ and $J_1$ denote Bessel functions of order zero and one. 

Thus the signal we are measuring is proportional to $({\mathcal F} \delta L/\lambda)$, as opposed to simply $(\delta L/\lambda)$ for the conventional Michelson interferometer. The finesse ${\mathcal F}$ of the cavity, and hence the enhancement to the sensitivity to the photothermal displacement $\delta L$, can, in principle, be made quite large, with values of ${\mathcal F} = 10^5$ achievable with standard techniques.

In addition to enhancing the sensitivity to a given photothermal displacement $\delta L$, the Fabry-Perot cavity also enhances the displacement itself. If $P_{pump}$ is the power in the pump beam, incident on the cavity, then the power incident on the sample inside the cavity is approximately $P_{inc} \approx {\mathcal F} P_{pump}$. As we shall see below, the photothermal response $\delta L$ is proportional to the total power absorbed by the sample, which in turn is proportional to $P_{inc}$. Thus the Fabry-Perot cavity enhances the signal measured at the lock-in amplifier by a factor of ${\mathcal F}^2$. 

This enhancement of the signal over conventional methods opens up many new possibilities for photothermal measurements. For example, it should be possible to measure the thermal properties of very-low-absorption samples that would otherwise be inaccessible to photothermal techniques. 

The sensitivity of this method is, in practice, limited by two things. First, it is limited by a combination of noise in the cavity length and laser-frequency noise. Intrinsic cavity length noise is thermal in origin and is typically small enough to be negligible for photothermal experiments except at the mechanical resonant frequencies of the cavity itself~\cite{Saulson90}. Extrinsic noise in the cavity can be suppressed by suitable isolation methods, including suspension from seismic isolation platforms and enclosure in a vacuum apparatus.

A solid-state Nd:YAG laser, such as the Lightwave 126~\cite{Lightwave}, typically has a frequency noise of $\delta \nu (f) \approx 100 Hz/\sqrt{Hz} \times (100 Hz/f)$, where $f$ is the measurement frequency~\cite{Abramovici96}. For a 30cm cavity, this corresponds to an equivalent length noise of 
\begin{eqnarray*}
\delta L_{eq-freq} (f) & = & \frac{L \lambda}{c} \delta \nu (f) \\
      & \approx  & 1 \times 10^{-13} \frac{m}{\sqrt{Hz}} \times  \left[ \frac{100 Hz}{f} \right] \left( \frac{L}{30cm} \right) \left( \frac{\lambda}{1.064 \mu m} \right).
\end{eqnarray*}

The fundamental shot noise limit for measuring the length of the cavity is much lower and, well below the cavity pole, is~\cite{Black01}
\begin{eqnarray*}
\delta L_{eq-shot} & = & \frac{\sqrt{hc}}{8} \frac{\sqrt{\lambda}}{\finesse \sqrt{P_{probe}}} \\
      & \approx &  8.1 \times 10^{-19} \frac{m}{\sqrt{Hz}} \times \left( \frac{100}{{\mathcal F}}  \right) \left( \frac{\lambda}{1.064 \mu m} \right)^{\frac{1}{2}} \left( \frac{500 mW}{P_{probe}} \right)^{\frac{1}{2}}.
\end{eqnarray*}

Second, the sensitivity is limited by the degree to which the pump and probe beams can be made orthogonal, and therefore isolated from each other. For the photothermal signals we have observed, we have had no difficulty in reducing this pump-probe cross-coupling noise to much less than the amplitude of our signal, even at the highest modulation frequencies and lowest signal levels.

\section{The photothermal response}
\label{sec:data-analysis}

A rigorous modeling of the photothermal response requires a complete, multidimensional solution to the diffusion equation to solve for the temperature distribution inside the sample, then a solution to the Navier-Stokes equations to solve for the resulting thermal expansion. This has been done in a variety of contexts~\cite{Balageas86, Li91, Liu94, Liu93, Vicanek94}, but the form of the solutions is somewhat involved, and fitting them to experimental data to extract a sample's thermal properties, especially those of a sample that includes a thin, inhomogeneous surface layer, is computationally intensive. 

We may gain considerable insight into the relationship between the photothermal response and thermal properties of a sample by making a simple estimate of the frequency dependence and magnitude of the photothermal response. If we apply a sinusoidal heat source to the surface of a material, thermal waves will propagate into the material, with a resulting sinusoidal thermal expansion over the volume being heated. A very general property of thermal waves is that they decay away as they propagate, with a characteristic decay length that is equal to the wavelength of the thermal wave. In this sense, they behave very much like electromagnetic waves propagating into a normal conductor~\cite{Jackson}. For a homogeneous material with thermal conductivity $\kappa$, mass density $\rho$, and specific heat $C$, this penetration depth $\ell_t$ is given by
\[
\ell_t (f) = \sqrt{\frac{\kappa}{\rho C} \frac{1}{\pi f}}.
\]
Thus we may assume, for a rough approximation, that the material only gets heated to a depth of $\ell_t$. The thermal expansion of this part of the material is given, again approximately, by
\[
\delta L = \ell_t \alpha \Delta T.
\]

Let's consider a sinusoidally modulated pump beam with a power given by
\[
P_{pump} = P_0 + P_m \frac{\sin ( 2 \pi f t )}{2} ,
\]
where $P_m$ is the peak-to-peak modulated power in the pump beam. (For $ 100 \% $ modulation, $P_0=P_m/2$, and $P_m$ is the total amplitude of the pump beam.) If $P_{abs}=W P_m$ is the (peak-to-peak) power absorbed at the surface of the sample ($W$ being the absorption coefficient), then the total temperature change $\Delta T$ over a single cycle is $\Delta T = (P_{abs}/2f)/(\rho C V)$, where $V$ is the volume of the material getting heated.

For high frequencies, where $\ell_t \ll r_0$ and $r_0$ is the radius of the laser spot doing the heating, we can approximate the volume by~\cite{Braginsky99}
\[
V \sim \pi r_0^2 \ell_t,
\]
which gives an approximate photothermal response
\[
\delta L_{hi-f} (f) \sim  P_{abs} \frac{\alpha}{\rho C} \frac{1}{r_0^2} \frac{1}{f},
\]
or
\begin{equation}
\label{eq:uncoating}
\delta L_{hi-f} (f) \sim P_{abs} \frac{\alpha}{\kappa} \frac{f_s}{f},
\end{equation}
where the characteristic frequency for this homogeneous substrate is defined as 
\begin{equation}
\label{eq:substrate-frequency}
f_s \equiv \frac{\kappa }{ \pi \rho C r_0^2}.
\end{equation}
The high-frequency photothermal response is proportional to $1/f$, and its magnitude gives us the thermal expansion coefficient $\alpha$, provided we know $P_{abs}$, $\rho$, and $C$. (Or it can give $P_{abs}$ if we know $\alpha$, $\rho$, and $C$, etc.)

For low frequencies, we approximate the volume by~\cite{Cerdonio01}
\[
V \sim \frac{1}{2} \left( \frac{4}{3} \pi \ell_t^3 \right),
\]
which gives a photothermal response of
\begin{equation}
\label{eq:low-f}
\delta L_{low-f} (f) \sim P_{abs} \frac{\alpha}{\kappa}.
\end{equation}
Note that the low-frequency response is essentially independent of frequency, with a natural ``turning point'' between the low- and high-frequency regimes at the crossover frequency $f_s$. This turning point does not depend on $P_{abs}$ or $\alpha$, is determined by the length scale $r_0$, and can be used to get the thermal conductivity $\kappa$, provided we know $r_0$ and $\rho C$.

A careful derivation gives the complete response, valid for a homogeneous material at all frequencies~\cite{Cerdonio01, DeRosa02}, as
\begin{equation}
\label{eq:all-f}
\delta L (f) = \left( \frac{P_{abs}}{2} \right) \frac{1}{ \pi} \frac{\alpha (1 + \sigma) }{\kappa} 
	\left|  \frac{1}{ \pi} \int_0^{\infty} du \int_{-\infty}^{+\infty} dv \frac{u^2 e^{-u^2/2}}{(u^2 + v^2) (u^2 + v^2 + i 2 f/f_s)} \right|
\end{equation}
where $\sigma$ denotes Poisson's ratio.



For a material with a coating, there is an additional length scale, the coating thickness $t$. At sufficiently high frequencies the thermal penetration depth $\ell_t$ becomes less than $t$, and the thermal waves essentially only sample the coating. When this happens, the photothermal response will be dominated by the coating and we have, approximately
\[
\delta L_{f \gg f_t} (f) \sim P_{abs} \frac{\alpha_c}{\kappa_c} \frac{f_c}{f},
\]
where the subscript $c$ denotes a coating property. The transition frequency that defines this regime occurs where $\ell_t = t$, or
\begin{equation}
\label{eq:transition-frequency}
f_t \equiv \frac{\kappa_c }{ \pi \rho_c C_c t^2},
\end{equation}
and the coating characteristic frequency is
\begin{equation}
\label{eq:coating-frequency}
f_c \equiv \frac{\kappa_c }{ \pi \rho_c C_c r_0^2}.
\end{equation}
At low frequencies, where $\ell_t \gg t$, the volume of the sample that gets heated and expands is predominantly substrate, with the coating contributing very little to the overall photothermal response. In this regime we expect the net response of the coated sample to be essentially indistinguishable from that of an uncoated sample.

\begin{center}
\begin{figure}
\includegraphics{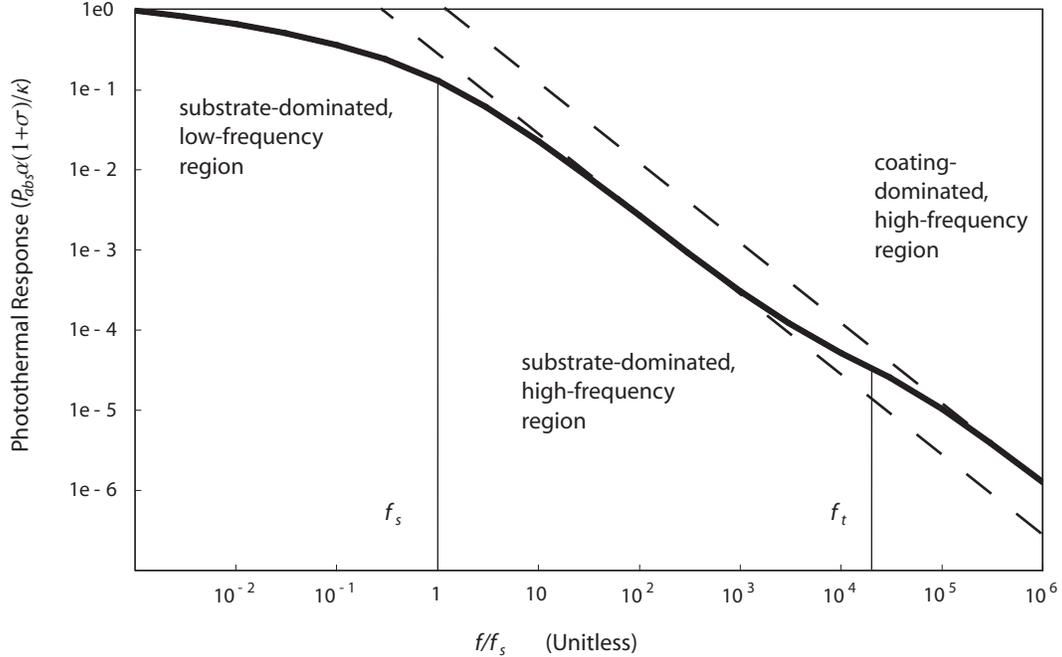}
\caption{\label{fig:approximation-plot} The expected photothermal response of a coated sample. When the coating thickness is much less than the laser spot radius, there are three clearly defined frequency regimes. The magnitude of the photothermal response in each regime and the transition frequencies between regimes together allow us to determine the thermal properties of the sample.}
\end{figure}
\end{center}

If $t \ll r_0$ there will be two high-frequency regimes, as shown in Figure~\ref{fig:approximation-plot}: one at moderately-high frequencies in which $t \ll \ell_t \ll r_0$, where the photothermal response has a $1/f$ frequency response and is dominated by the substrate, and another at higher frequencies where $\ell_t  \ll t  \ll r_0$. In this highest-frequency regime, the frequency dependence is again $1/f$, but the photothermal response is now dominated by the coating. The crossover frequency between these two regimes occurs the frequency where $\ell_t = t$, or $f = f_t$.

We may interpolate between these two high-frequency regimes to obtain an approximate, high-frequency photothermal response of a coated sample.
\begin{equation}
\label{eq:coating}
\delta L (f) \approx \frac{P_{abs}}{4 \pi^2} \left[ \left( \frac{\alpha (1 + \sigma)}{\kappa} \right) \frac{f_s}{f} + \left( \frac{\alpha_c (1 + \sigma_c)}{\kappa_c} \right) \frac{f_c}{f_t + f} \right]
\end{equation}
This is the formula we will use to fit our high-frequency data to extract the coating thermal properties.

\section{Results}

We measured the photothermal response of two samples. The first was single-crystal, synthetic, c-axis sapphire. The second was identical to the first except for the addition of a $4 \mu m$ multilayer dielectric coating of alternating layers of $\text{SiO}_2$ and $\text{TiO}_2$. This coating formed a high-reflectivity mirror for infrared light with a wavelength of $1.064 \mu m$~\cite{Rao03}. Both samples, including the dielectric coating, were provided by CVI Laser, Inc.~\cite{CVI}

Both samples had a thin ($200 nm$) layer of gold deposited on their surfaces. The purpose of this gold layer was threefold. First, this matched the optical reflectivities of the two samples so that all other aspects of the measurements would be the same. Second, the gold boosted the absorption for these initial measurements to enhance the photothermal response signal. And third, we wanted to ensure that all of the absorption occurred in the gold layer at the surfaces of the samples, as opposed to somewhere in the bulk or deep within the coating. Using the tabulated values of the thermal expansion coefficient, density, and specific heat for bulk gold~\cite{CRC}, and the reported thermal conductivity for a $200 nm$ gold film~\cite{Wu93}, we estimate that because the thermal conductivity of the gold layer is so good, and its thickness so small, its photothermal response will be negligible below $250 MHz$. At those frequencies, the observed response will be dominated by the substrate and dielectric coating. 




Agreement between theory and experimental data is excellent at all frequencies, giving us high confidence in the measurement techniques. Moreover, the values for $\alpha (1+\sigma)$ and $\kappa$ are consistent with our expectations from tabulated bulk values. (See Table~1.)

\begin{center}
\begin{figure}
\includegraphics{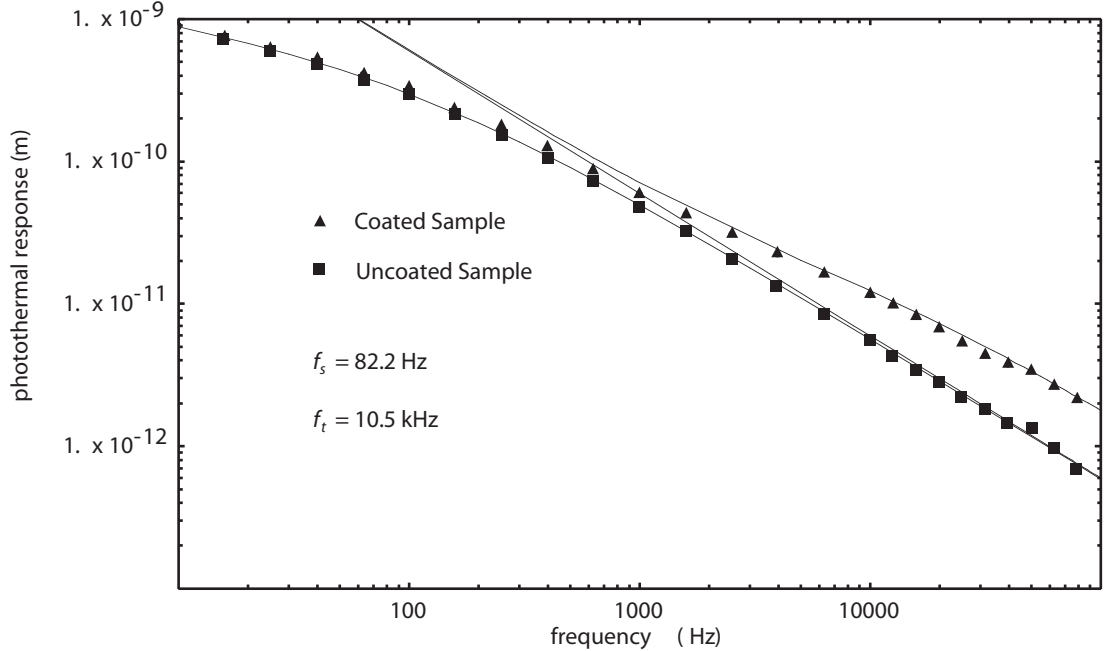}
\caption{\label{fig:high-freq} Photothermal response of both samples. Upper dots are experimental data for Sample 2, with the dielectric coating. The lowest theory curve is that of Reference~\cite{Cerdonio01}, which accounts for the substrate's photothermal response at all frequencies. The other two curves show high-frequency asymptotic behavior for samples with (Equation~\ref{eq:coating}) and without (Equation~\ref{eq:all-f}) dielectric coatings. Noise in the data at high frequencies (low amplitudes) is due to optical crosstalk between the pump and probe beams. These effects are systematic and appear at the same levels and with the same frequency dependence in both samples. Gaussian error bars are smaller than the points as illustrated.}
\end{figure}
\end{center}

Figure~\ref{fig:high-freq} shows the photothermal response of both samples, where the upper set of data points are from the sample with the dielectric coating. The thermal properties used in this fit are given in Table~1. The two parameters we varied for the fit were the thermal conductivity $\kappa$, which determined the rolloff frequency, and the product $\alpha (1 + \sigma)$, which determined the overall amplitude. The bulk value for the product $\rho C$ was assumed for both coating and substrate, and the absorbed power was measured from the visibility of the cavity. We measure the sum of the transmitted and reflected power and subtract it from the incident power to find the total power lost in the cavity. We then assume that this loss is due to absorption, rather than scattering. This assumption is only justified because of the relatively high-absorption gold film present in our sample. The response of the coated sample was essentially identical to that of the uncoated sample at low frequencies, which agrees well with our expectations that the dominant photothermal response in that regime is due to the substrate. This agreement also gives us confidence that the absorption is the same for the two samples, which was one of the goals in applying the thin gold coatings. 

At high frequencies, the response of the coated sample differs substantially from that of the uncoated one, and it agrees well with our expectations based on the thermal penetration depth as outlined in Section~\ref{sec:data-analysis}. Using the tabulated values of $\rho$ and $C$ for bulk $\text{SiO}_2$ and $\text{TiO}_2$~\cite{CRC}, we extract an effective thermal expansion coefficient and thermal conductivity of $\alpha_c ( 1 + \sigma_c ) = (8.6 \pm 0.6) \times 10^{-6} K^{-1}$ and $\kappa_c = (1.08 \pm 0.15) W/mK$. This value of the thermal conductivity is less than would be expected from the bulk values of $\kappa$ for $\text{SiO}_2$ and $\text{TiO}_2$, which are $1.18 W/mK$ and $10.4 W/mK$, respectively, which together would give an effective thermal conductivity of $1.83 W/mK$. This reduction of the thermal conductivity is consistent with other observations, which indicate that the effective thermal conductivity of a material in thin-film form is often lower that its bulk value~\cite{Wu93,Grilli00,Wu97,Kuo92,Langer97}. 

\begin{center}
\begin{tabular}{|| l | c | c | c | c ||} \hline
Material	&	$\rho\ (kg/m^3)$	&	$C\ (J/kg K)$	&	$\alpha\ (K^{-1})$	&	$\kappa\ (W/m K)$	\\ \hline
$\text{SiO}_2$	&	$2.2\times10^3$~\cite{Braginsky03}	&	$670$~\cite{Braginsky03}	&	$5.5\times10^{-7}$~\cite{Braginsky03}	&	$1.4$~\cite{Braginsky03}	\\
$\text{TiO}_2$	&	$4.23\times10^3$~\cite{CRC}	&	$688$~\cite{CRC}	&	$5\times10^{-5}$~\cite{Braginsky03}	&	$10.4$~\cite{Wu93}	\\
Gold film	&	$19.3\times10^3$	&	$126$ &	 &	$25$~\cite{Wu93}	\\ \hline
	&		&		&	$\alpha(1+\sigma)\ (K^{-1})$	&		\\ \hline
Sapphire	&	$4.0\times10^3$~\cite{Braginsky03}	&	$790$~\cite{Braginsky03}	&	$(5.53 \pm 0.06)\times10^{-6}$*	&	$64 \pm 2*$	\\
Coating	&	$3\times10^3$	&	$680$	&	$(8.6 \pm 0.6)\times10^{-6}*$	&	$1.08\pm0.15*$	\\ \hline
\end{tabular}

{\bf Table 1:} Relevant material properties used in this work. Entries with an asterisk (*) denote values derived from our data.
\end{center}

\section{Conclusions}

We have developed a new method of interferometric photothermal displacement spectroscopy that is well-suited for studying thin dielectric films of the type used in optical coatings. Our method uses a Fabry-Perot cavity to enhance both the sensitivity of the interferometric measurement and the amplitude of the photothermal response (by enhancing the pump-beam power). Together, these can potentially provide orders of magnitude more sensitivity than conventional photothermal displacement methods. In this paper we demonstrate both the technique and the data analysis by measuring the thermal conductivity and thermal expansion coefficient of a high-reflectivity $\text{SiO}_2-\text{TiO}_2$ film on a sapphire substrate. 

\section{Acknowledgments} 

Many thanks to Alan Weinstein for carefully reading this manuscript and for providing many helpful suggestions. This work was supported by the NSF under grant number PHY98-01158.

\newpage

\bibliographystyle{unsrt}
\bibliography{Photothermal}

\begin{thebibliography}{10}

\bibitem{Fejer03}
M.~M. Fejer, S.~Rowan, G.~Cagnoli, D.~R.~M. Crooks, A.~Gretarsson, G.~M. Harry,
  J.~Hough, S.~D. Penn, P.~H. Sneddon, and S.~P. Vyatchanin.
\newblock Thermoelastic dissipation in inhomogeneous media: loss measurements
  and displacement noise in coated test masses for interferometric
  gravitational wave detectors.
\newblock {\em Physical Review D}, 2003.
\newblock Submitted to Physical Review D.

\bibitem{Braginsky03}
V.~B. Braginsky and S.~P. Vyatchanin.
\newblock Thermodynamical fluctuations in optical mirror coatings.
\newblock {\em Physics Letters A}, 312(3-4):244--255, 2003.

\bibitem{Braginsky00}
V.~B. Braginsky, M.~L. Gorodetsky, and S.~P. Vyatchanin.
\newblock Thermo-refractive noise in gravitational wave antennae.
\newblock {\em Physics Letters A}, 271:303--307, 2000.

\bibitem{Nakagawa02}
N.~Nakagawa, A.~M. Gretarsson, and E.~K. Gustafson.
\newblock Thermal noise in half infinite mirrors with non-uniform loss: a slab
  of excess loss in a half-infinite mirror.
\newblock {\em Physical Review D}, 65(10):102001, 2002.

\bibitem{Yamamoto02-1}
K.~Yamamoto, M.~Ando, and K.~Kawabe.
\newblock Thermal noise caused by an inhomogeneous loss in the mirrors used in
  the gravitational wave detector.
\newblock {\em Physics Letters A}, 305:18--25, 2002.

\bibitem{Yamamoto02-2}
K.~Yamamoto, S.~Otsuka, and M.~Ando.
\newblock Study of the thermal noise caused by inhomogeneously distributed
  loss.
\newblock {\em Class. Quantum Grav.}, 19:1689--1696, 2002.

\bibitem{Harry02}
G.~M. Harry, A.~M. Gretarsson, P.~R. Saulson, S.~E. Kittelberger, S.~D. Penn,
  W.~J. Startin, S.~Rowan, M.~M. Fejer, D.~R.~M. Crooks, G.~Cagnoli, J.~Hough,
  and N.~Nakagawa.
\newblock Thermal noise in interferometric gravitational wave detectors due to
  dielectric coatings.
\newblock {\em Class. Quantum Grav.}, 19:897--917, 2002.

\bibitem{Crooks02}
D.~R.~M. Crooks, P.~Sneddon, G.~Cagnoli, J.~Hough, S.~Rowan, M.~M. Fejer,
  E.~Gustafson, R.~Route, N.~Nakagawa, D.~Coyne, G.~M. Harry, and A.~M.
  Gretarsson.
\newblock Excess mechanical loss associated with dielectric mirror coatings on
  test masses in interferometric gravitational wave detectors.
\newblock {\em Class. Quantum Grav.}, 19(5):883--896, 2002.

\bibitem{Braginsky99}
V.~B. Braginsky, M.~L. Gorodetsky, and S.~P. Vyatchanin.
\newblock Thermodynamical fluctuations and photo-thermal shot noise in
  gravitational wave antennae.
\newblock {\em Physics Letters A}, 264:1--10, 1999.

\bibitem{Cerdonio01}
M.~Cerdonio, L.~Conti, A.~Heidmann, and M.~Pinard.
\newblock Thermoelastic effects at low temperatures and quantum limits in
  displacement measurements.
\newblock {\em Physical Review D}, 63:082003-- 1--9, 2001.

\bibitem{Wu93}
Z.~L. Wu, P.~K. Kuo, Lanhua Wei, S.~L. Gu, and R.~L. Thomas.
\newblock Photothermal characterization of optical thin films.
\newblock {\em Thin Solid Films}, 236:191--198, 1993.

\bibitem{Grilli00}
M.~L. Grilli, D.~Ristau, M.~Dieckmann, and U.~Willamowski.
\newblock Thermal conductivity of e-beam coatings.
\newblock {\em Applied Physics A}, 71:71--76, 2000.

\bibitem{Wu97}
Z.~L. Wu, M.~Thomsen, and P.~K. Kuo.
\newblock Photothermal characterization of optical thin film coatings.
\newblock {\em Opt. Eng.}, 36(1):251--262, 1997.

\bibitem{Kuo92}
B.~S.~W. Kuo, J.~C.~M. Li, and A.~W. Schmid.
\newblock Thermal conductivity and interface thermal resistance of {Si} film on
  {Si} substrate determined by photothermal displacement interferometry.
\newblock {\em Applied Physics A}, 55:289--296, 1992.

\bibitem{Langer97}
G.~Langer, J.~Hartmann, and M.~Reichling.
\newblock Thermal conductivity of thin metallic films measured by photothermal
  profile analysis.
\newblock {\em Review of Scientific Instruments}, 68(3):1510--1513, 1997.

\bibitem{Braginsky03-2}
V.~B. Braginsky and A.~A. Samiolenko.
\newblock Measurement of the optical mirror coating properties.
\newblock pages gr--gc/0304100 v1, 2003.

\bibitem{Inci02}
M.~N. Inci.
\newblock Simultaneous measurements of thermal optical and linear thermal
  expansion coefficients of {Ta2O5} films.
\newblock In {\em ICO 19}, Firenze, Italy, 2002.

\bibitem{Tien00}
C.~L. Tien.
\newblock Simultaneous determination of the thermal expansion coefficient and
  the elastic modulus of {Ta2O5} thin films using phase shifting
  interferometry.
\newblock {\em J. of Mod. Opt.}, 47(10):1681--1691, 2000.

\bibitem{Olmstead83}
M.~A. Olmstead, N.~M. Amer, S.~Kohn, D.~Fournier, and A.~C. Boccara.
\newblock Photothermal displacement spectroscopy: An optical probe for solids
  and surfaces.
\newblock {\em Applied Physics A}, 32:141--154, 1983.

\bibitem{Mandelis-books-1}
A.~Mandelis, editor.
\newblock {\em Progress in photothermal and photoacoustic science and
  technology}, volume~1.
\newblock Elsevier, New York, 1992.

\bibitem{Mandelis-books-2}
A.~Mandelis, editor.
\newblock {\em Progress in photothermal and photoacoustic science and
  technology}, volume~2.
\newblock Prentice-Hall, Englewood Cliffs, N. J., 1994.

\bibitem{Mandelis-books-3}
A.~Mandelis and P.~Hess, editors.
\newblock {\em Progress in photothermal and photoacoustic science and
  technology}, volume~3.
\newblock SPIE Opt. Eng. Press, Bellingham, Wash., 1997.

\bibitem{Mandelis-books-4}
A.~Mandelis and P.~Hess, editors.
\newblock {\em Progress in photothermal and photoacoustic science and
  technology}, volume~4.
\newblock Bellingham, Wash., New York, 2000.

\bibitem{Mandelis00}
Andreas Mandelis.
\newblock Diffusion waves and their uses.
\newblock {\em Physics Today}, 53(8):29--34, 2000.

\bibitem{DeRosa02}
M.~DeRosa, L.~Conti, M.~Cerdonio, M.~Pinard, and F.~Marin.
\newblock Experimental measurement of photothermal effect in {F}abry-{P}erot
  cavities.
\newblock {\em Physical Review Letters}, 89(23), 2002.

\bibitem{Black00}
E.~D. Black.
\newblock How to measure {B}raginsky's photo-thermal effect.
\newblock {\em LIGO Technical Document}, T000001-00-R:1--10, 2000.

\bibitem{Rao03}
Shanti~R. Rao.
\newblock {\em Mirror Thermal Noise in Interferometric Gravitational Wave
  Detectors}.
\newblock PhD thesis, California Institute of Technology, 2003.

\bibitem{Drever83}
R.~W.~P. Drever, J.~L. Hall, F.~V. Kowalski, J.~Hough, G.~M. Ford, A.~J.
  Munley, and H.~Ward.
\newblock Laser phase and frequency stabilization using an optical resonator.
\newblock {\em Appl. Phys. B: Photophys. Laser Chem.}, 31:97--105, 1983.

\bibitem{Black01}
E.~D. Black.
\newblock An introduction to {P}ound-{D}rever-{H}all laser frequency
  stabilization.
\newblock {\em Am. J. Phys.}, 69(1):79--87, 2001.

\bibitem{Saulson90}
P.~R. Saulson.
\newblock Thermal noise in mechanical experiments.
\newblock {\em Physical Review D}, 42(8):2437--2445, 1990.

\bibitem{Lightwave}
Lightwave Electronics.
\newblock 2400 {C}harleston {R}oad, {M}ountain {V}iew {CA} 94043.
\newblock http://www.lightwaveelectronics.com.

\bibitem{Abramovici96}
A.~Abramovici and R.~Savage.
\newblock {NPRO-PSL Conceptual Design}.
\newblock {\em LIGO Technical Document}, pages T960089--00--D, 1996.

\bibitem{Balageas86}
D.~L. Balageas, J.~C. Krapez, and P.~Cielo.
\newblock Pulsed photothermal modeling of layered materials.
\newblock {\em Journal of Applied Physics}, 59(2):348--357, 1986.

\bibitem{Li91}
Bingcheng Li, Zhaoxin Zhen, and Shunhua He.
\newblock Modulated photothermal deformation in solids.
\newblock {\em Journal of Physics D: Applied Physics}, 24:2196--2201, 1991.

\bibitem{Liu94}
M.~Liu, M.~B. Suddendorf, and M.~G. Somekh.
\newblock Response of interferometer based probe systems to photodisplacement
  in layered media.
\newblock {\em Journal of Applied Physics}, 76(1):207--215, 1994.

\bibitem{Liu93}
M.~Liu, M.~B. Suddendorf, and M.~G. Somekh.
\newblock {\em Semicond. Sci. Technol.}, 8:1639, 1993.

\bibitem{Vicanek94}
M.~Vicanek, A.~Rosch, F.~Piron, and G.~Simon.
\newblock Thermal deformation of a solid surface under laser irradiation.
\newblock {\em Applied Physics A}, 59:407--412, 1994.

\bibitem{Jackson}
J.~D. Jackson.
\newblock {C}lassical {E}lectrodynamics.
\newblock pages 296--298. John Wiley and Sons, New York, 2 edition, 1975.

\bibitem{CVI}
{CVI Laser, LLC}, 200 Dorado Place SE, Albuquerque, NM 87123.

\bibitem{CRC}
David~R. Lide, editor.
\newblock {\em CRC Handbook of Chemistry and Physics}.
\newblock CRC Press, Boca Raton, 77 edition, 1996-1997.

\end{thebibliography}

\end{document}